\documentclass[10pt, twocolumn]{IEEEtran}

\UseRawInputEncoding

\usepackage{graphicx}

\usepackage{amsopn}
\usepackage{multirow}
\usepackage{amsxtra,amssymb}
\usepackage[english]{babel}
\ifCLASSINFOpdf
\else
\fi

\begin{document}

\title{Backscatter Communication System With Dumb Diffusing Surface}

%

\author{
\IEEEauthorblockN{
Jean-Marc Kelif and Dinh-Thuy Phan-Huy\\
Orange Labs Networks, \\
Chatillon, France. \\
E-mail: jeanmarc.kelif@orange.com, dinhthuy.phanhuy@orange.com \\} 
}



%


\maketitle

\begin{abstract}
Ambient backscatter communications have been identified for ultra-low energy wireless communications. Indeed, a tag can send a message to a reader without emitting any wave and without battery, simply by backscattering the waves generated by a source. In the simplest implementation of such a system, the tag sends a binary message by oscillating between two states and the reader detects the bits by comparing the two distinct received powers. In this paper, for the first time, we propose to study an ambient backscatter communication system, in the presence of a diffusing surface, a simple flat panel that diffuses in all directions. We establish the analytical closed form expression of the power contrast in the presence of the surface. We show that the diffusing surface improves the power contrast. 
Moreover our approach allows us to express the contrast to noise ratio, and therefore to establish the BER performance. 
Furthermore, we derive the optimum source transmit power for a given target power contrast. 
This makes it possible to quantify the amount of energy that can be saved at the source side, thanks to the diffusing
surface.

\end{abstract}


\IEEEpeerreviewmaketitle



%

\section{Introduction}

Ambient backscatter communication systems \cite{Liu13} have been identified as a promising technology for ultra-low energy wireless communications. In such a communication system, as illustrated in Figure 1, a tag T transmits a message to a reader R without generating any radio wave and without battery. Indeed, T reflects, in a variable manner in time, the waves generated by a source S. S can be a DVB-T tower or a Wi-Fi access point \cite{Huynh18}. It could also be a 4G base station \cite{RachediPhan19}, a 5G base station or device \cite{Fara20} \cite{Gati19}. In its most general form \cite{Huynh18}, T is a resonant antenna in the frequency band of S connected to a variable load impedance. The load impedance is variable and takes several distinct values. Each impedance corresponds to a specific data symbol, and a specific backscatter state of T.\\

T emits a message by modulating its impedance. R detects variations in the signal, backscattered by T, and demodulates the received signal. In its simplest implementation \cite{Liu13}, T is a dipole connected to a switch which puts the dipole either in open circuit (infinite impedance) or in closed circuit (zero impedance), and two distinct symbols are emitted. The simplest detector is an energy detector \cite{Wan16} to demodulate a simple differential encoding of data 
whose performance increases with the power contrast between the two states \cite{Wan16}. However, the performance of such a detector is limited by the power contrast between the states \cite{Yan18}. \\
Recent works improve the detection of T by R, either by using a more sophisticated detector with Channel State Information at R side, that exploits pilots sent by S as in \cite{Devi19}, or by synchronizing the transmission of Orthogonal Frequency Division Multipex (OFDM) symbols by S and with the switching of T, so that T switches always occur in the middle of an OFDM symbol \cite{Yan18}.

Recently, it was proposed in \cite{ITU15} to improve radio links using large reconfigurable intelligent meta-surfaces in the environment. such a material is in fact set of discrete elements. \\
It is an array of reconfigurable and electronically tunable reflectors. Deployed in the environment, on walls or panels, and optimally tuned, such a surface can direct an incident wave into a target direction. They could be used to improve the power contrast, and the detection of the tag by the reader.
However intelligent surfaces are complex and costly.\\

For the first time, we propose to investigate and quantify whether a dumb and passive simple diffusing surface deployed in the environment (on walls for instance) could already improve the performance of an ambient backscatter communication system. We propose to use, a planar and continuous surface of a rough material (compared to the wavelength of the source radio waves), such as the one described in \cite{ITU15} paragraph 2.3.2, for instance. In this example, the surface is flat in average and has cylindrical asperities whose radii are of the order of the wavelength.\\

Such a material has the particularity to backscatter radio waves in every direction. Contrary to \cite{Renzo19}, such a diffusive surface is not intelligent and cannot be electronically reconfigured to change its backscattering or reflecting properties. However, it is less complex and less costly. More precisely, our analysis involves an ambient backscatter communication system deployed close to a diffusive surface. The considered tag is the most simple. It oscillates between two states: transparent and backscattering states, respectively \cite{Wan16}. The considered reader uses the most simple receiver technique: the energy detector. As the reader measures the power received in the two states, its performance increases with the power contrast $\Delta P_s$. As illustrated in Figure 2, we consider a deployment scenario where the surface is perpendicular to the plane containing the source, the transmitter and the reader. This would be typically the case if the surface were deployed on a wall. Figure 2 illustrates the waves propagation between the source S and the reader R, for the two states of the tag: transparent state tag, in Figure~2-a), and retro-diffusing state tag, in Figure~2-b).\\

\textbf{Our first contribution:} we develop an analytical model of the aforementioned ambient backscatter communications system deployed near a dumb diffusing surface. This model establishes a closed form expression of the power received by the reader, in the two states of the tag, transparent or retro-diffusing. To the best of our knowledge, this is the first time such a formula is established. Indeed, many exact bit error rate analysis with various detectors have been derived for a propagation channel environment with Rayleigh fading \cite{Wan16} \cite{Yan18} \cite{Devi19} \cite{Devi18}, i.e. with scatterers randomly distributed in space around the tag and the reader. However, the impact of a dumb diffusing surface has not been studied yet. \\

\textbf{Our second contribution:} thanks to the closed form expression of $\Delta P_s$, we can derive the contrast to noise ratio reached at the reader, which allows us to evaluate the BER performance of the system. Moreover, we show that the presence of the diffusing surface improves the detection of the
message transmitted by the tag, by increasing the power contrast, in many configurations of the system.

\begin{figure}[htbp]
\centering
\includegraphics[scale=0.25]{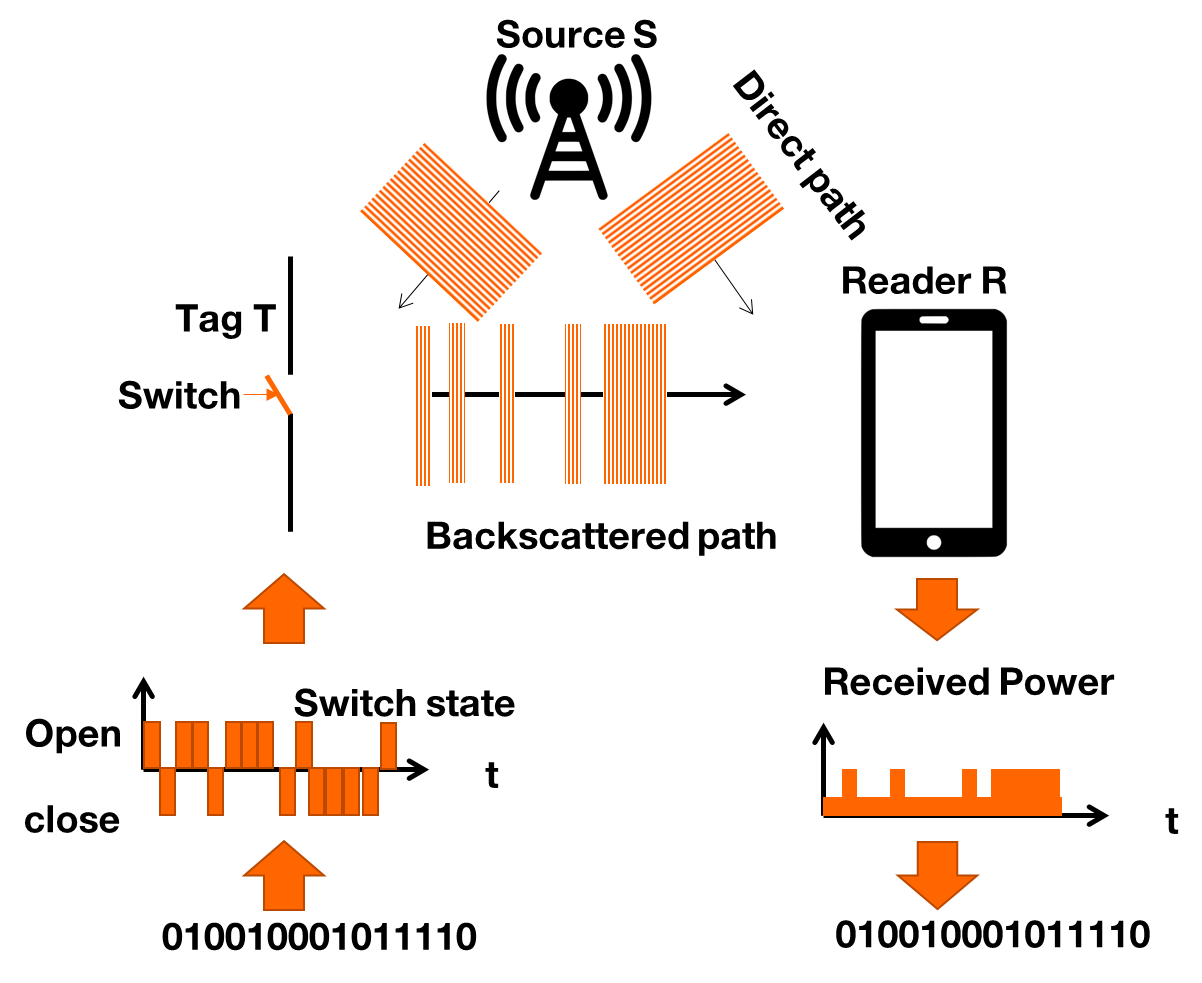}
\caption{\footnotesize State-of-the-art ambient backscatter communication.}
\label{StateArt}
\end{figure}


\begin{figure}[htbp]
\centering
\includegraphics[scale=0.5]{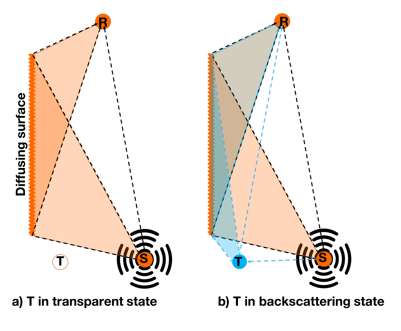}
\caption{\footnotesize Ambient backscatter communication aided by a diffusing surface.}
\label{Ambientbackscatterdiffusingsurface}
\end{figure}



The organization of the paper follows. 
In Section \ref{systemparametersnotations}, the system parameters and notations are defined. 
In Section \ref{systemmodel}, we present the system model and establish the closed form formula of the power contrast by considering a diffusing surface. 
In Section \ref{sec:powercontrast}, we propose an analysis of the power contrast impact. 
Section \ref{sec:application} focuses on applications of the closed form formula.
Section \ref{sec:conclusion} concludes the paper.

\section{System Parameters} \label{systemparametersnotations}

\begin{figure}[htbp]
\centering
\includegraphics[scale=0.38]{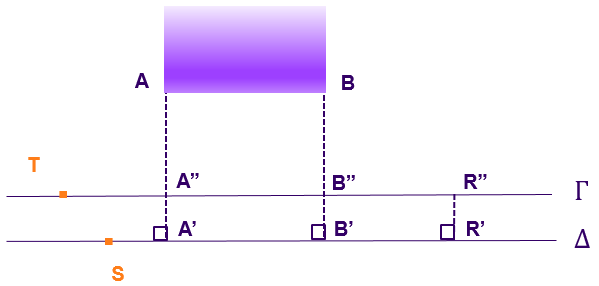}
\caption{\footnotesize System geometry}
\label{systemgeometry}
\end{figure}

The geometry of the system, constituted by the source S, the tag T, the diffusing surface AB and the reader R, is presented in Figure \ref{systemgeometry}. 
The following notations are considered \cite{Grffin09}.

\subsection{Backscatter Parameters} \label{systemparametersnotations2}
A power P is transmitted by a source located at S. The reader is located at R (Figure \ref{systemgeometry}).
A and B represent the extremities of the diffusing surface V.
$A', B'$ and $R'$ (resp. $A'', B''$ and $R''$) are the projections of A, B and R on ($\Delta$) (resp. ($\Gamma$)).
$G^s$ (resp. $G^r$ and $G^t$) is the antenna gain of the source (resp. the reader and the tag)
and $G^d$ is the diffusing coefficient of the surface V.
We denote $\lambda = \frac{c}{f}$ the wave length of the signal transmitted by the source, where c is the light speed, f the carrier frequency, k is the wave vector.
We denote $\omega$ =  $2 \pi f$, $K^{sr}$ =   $\frac{G^s G^r \lambda^2}{4 \pi}$, $K^{sar}$ = $\frac{G^s G^d G^r \lambda^4}{16 \pi^2}$,
and $K^{str}$ = $\frac{G^s (G^t)^2 G^r \lambda^4}{64 \pi^3}$ [7].

\subsection{Geometric parameters} \label{systemparametersnotations3}
The system model is presented in Figure \ref{model}.

We denote $D$ (resp. $a$ and $b$) the projection of the distance between the source S and the reader R on ($\Delta$) (resp. between A and S, and between B and R), and $d$ is the distance between the diffusing surface V and ($\Delta$).
We denote 
$h$ the length AB of the diffusing surface V, 
$r$ the distance SR (source-reader), 
$r_1$  the distance SA (source-extremity A of V), 
$r'_1$  the distance AR (extremity A of V-reader), 
$r_n$  the distance SB (source-extremity B of V), 
$r'_n$  the distance BR (extremity B of V-reader), 
$r_t$  the distance ST (source-tag),
$r'_t$ the distance TR (tag-reader), 
and $x$ the location of the received power at the reader.

\section{System Model} \label{systemmodel}

\begin{figure}[htbp]
\centering
\includegraphics[scale=0.62]{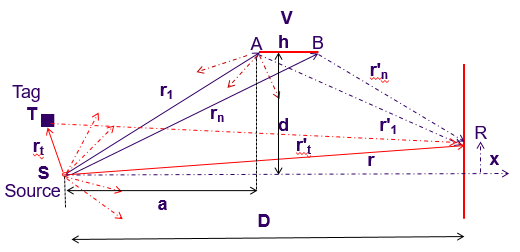}
\caption{\footnotesize System model.}
\label{model}
\end{figure}

Let consider a punctual source S emitting an isotropic electromagnetic wave. 
This wave is diffused by a tag $T$, in a retro-diffusing state, and a flat diffusing surface $V$ (Figure \ref{model}). The width of the surface is very small compared to its length, so that it can be considered negligible, and assumed to be one-dimensional. The reader $R$ receives multipath electromagnetic waves, constituted by the wave directly emitted by the source, the wave retro-diffused by the tag, and the wave diffused by the surface (Figure \ref{model}). \\
\textbf{Remark }:
Let notice that the amplitudes of multiple diffusions are negligible. Therefore a unique diffusion is considered, by the tag or by the diffusing surface, before arriving at the reader. Particularly, in the backscattering state, the signal diffused a first time by the tag, and diffused a second time by the surface, is not considered.


\subsection{Backscatter System} \label{sec:singlesource}

\subsubsection{Signal at the Reader} \label{sec:singlesource}
Considering the system (Figure \ref{model}) with a tag T in a retro-diffusing state, 
the signal $Y_{R}$ received by the reader can be expressed as 

\begin{eqnarray} \label{signaldirect1}
 Y_{R} &=& Y_{dir} + Y_{dif} + Y_{tag}	
\end{eqnarray}
where
\begin{eqnarray} \label{Pdirecte}
Y_{dir}=A_{dir} e^{i\omega t} e^{-ik r}
\end{eqnarray}
is the amplitude of the signal received by the reader, directly from the source,
and 
\begin{eqnarray} \label{Pdirecte}
Y_{tag}= A_{tag} e^{i\omega t}  e^{-ik(r_t+r'_t)}
\end{eqnarray}
is the amplitude of the signal emitted by the source S, retro-diffused by the tag T, and received by the reader R,
where 
\begin{eqnarray} \label{AdirAtag}
A_{dir} &=& \sqrt{\frac{K^{sr} P}{r^2}}, \nonumber \\	
A_{tag}&=&\sqrt{\frac{K^{str} P}{r_t^2 {r'}_t^2}}.
\end{eqnarray}

\subsubsection{Signal $Y_{dif}$ at the Reader} \label{sec:singlesource}
Let express the signal diffused by the surface V, received at a location $x$ of the reader, 
This signal is constituted by the set of all the signals diffused at each point of the surface.
Considering that $n$ rays coming from the source are diffused by V, this signal can be expressed as
\begin{eqnarray} \label{Ydifsum}
 Y_{dif} &=&   \sum_{j=1}^n \sqrt{\frac{K^{sar}P}{r_j^2 {r'}_j^2}}  \exp i(wt-kr_j)  \exp i(wt-k{r'}_j) \nonumber \\
				 &=&  \sum_{j=1}^n \sqrt{\frac{K^{sar}P}{r_j^2  {r'}_j^2  }} \exp i(wt-k(r_j+{r'}_j)) \nonumber \\
\end{eqnarray}
where $r_1...r_j...r_n$ represent the distances (between S and V) of the n rays coming from the source and reaching n points of the diffusing surface, and $r'_1...r'_j...r'_n$  represent the distances (between V and R) of the n waves diffused by the surface and reaching the point $x$ of the reader (Figure \ref{model}).

\subsection{Diffusing Surface Continuum Model} \label{sec:diffusingSurface}

The surface V is considered as a continuum. Therefore, the diffusion of rays coming from the source, is done by each infinitesimal element of V of length $du$.
Therefore  the signal emitted by the source S, diffused by V and received by the reader can be approximated as (see Appendix):
\begin{eqnarray} \label{Ydif1}
Y_{dif}=A_{dif} e^{i\omega t}  e^{-ik(r_1+r'_1)} \int_{V} e^{-ik u}du. 
\end{eqnarray}
From (\ref{Ydifsum}) and (\ref{Ydif1}), the amplitude of the signal at the reader, diffused by the surface V can be approximated as (see Appendix)
\begin{eqnarray} \label{signaldifcalcul}
 Y_{dif}  &=&  A_{dif} e^{i(wt)} e^{-ik(r_1+{r'}_1 +\frac{e}{2} )} \sin{\frac{k e}{2}},    \nonumber \\
\end{eqnarray}
where 
\begin{eqnarray} \label{Adif}
A_{dif} &=& \frac{2}{k} \sqrt{\frac{K^{sar}P}{r_1^2 {r'}_1^2}} 	
\end{eqnarray}
and denoting $e = h(\frac{a}{r_1} - \frac{b}{{r'}_1})$.

\subsection{Contrast Power Formula} \label{sec:diffusingSurface}

With a tag T in a retro-diffusing state, the total power of the signal received at a location $x$ (Figure \ref{model}) of the reader can be expressed as   

\begin{eqnarray} \label{signaldirect1}
 P_{R} &=&  Y_{R} Y_{R}^*  \nonumber \\	
				 &=& (Y_{dir} + Y_{dif} + Y_{tag} )(Y_{dir} + Y_{dif} + Y_{tag})^*. \nonumber \\	
\end{eqnarray}
From the established expressions of $Y_{dir}$, $Y_{dif}$ and $Y_{tag}$,
the power received by the reader can be expressed as
\begin{eqnarray} \label{signaldirect1}
 P_{R} &=& P_{1} +  \Delta P_s   
\end{eqnarray}
where 
\begin{eqnarray} \label{signaldirect1}
 P_1 &=&   P_{dir} + P_{dif} +  P_{dir-dif}
\end{eqnarray}
characterizes the power received at the reader for a tag in a transparent state.
We have
\begin{eqnarray} \label{Pdirecte}
P_{dir} &=& A_{dir}^2 \nonumber \\	
P_{dif}&=&  A_{dif}^2  \sin^2 (\frac{ke}{2})\nonumber \\	
P_{dir-dif}&=& 2 A_{dir} A_{dif}\sin⁡(\frac{ke}{2})\cos⁡[k(r_1+ r'_1-r+\frac{e}{2})] \nonumber \\
\end{eqnarray}
where $P_{dir}$ is the power due to the emitted source, $P_{dif}$ is the power diffused by the surface V, and 
$P_{dir-dif}$ characterizes the power due to an interferometry between direct and diffused waves.
And 
\begin{eqnarray} \label{deltaP1}
 \Delta P_s &=&  P_{tag} +   P_{dir-tag} +  P_{dif-tag} 
\end{eqnarray}
represents the power between the two states of the tag, with the presence of the diffusing surface.
where 
\begin{eqnarray} \label{deltaP3}
P_{tag} &=& A_{tag}^2 \nonumber \\
P_{dir-tag}  &=& 2 A_{dir} A_{tag} \cos⁡(k(r_t+r'_t-r)) \nonumber \\
P_{dif-tag}  &=& 2 A_{dif} A_{tag} \sin⁡(\frac{ke}{2}) \nonumber \\
							&\times& \cos⁡[k(r_1+ r'_1-(r_t+r'_t)+\frac{e}{2})]. \nonumber \\
\end{eqnarray}
In the case of a tag in a transparent state, $\Delta P_s =$0.

\section{Power Contrast} \label{sec:powercontrast}

The difference of powers received by R, depending on whether the tag T is in a backscattering or transparent state, makes it possible to characterize the efficiency of the system.
The expression of $\Delta P_s$ (\ref{deltaP1}) (\ref{deltaP3}) highlights the impact of
system parameters such as the location of the source,
the tag, the reader, the diffusing surface V, the wave transmission frequency.
In addition, performance analysis can be performed
quickly and easily, thanks to this expression.
Indeed, this formula makes it possible to quantify with accuracy the impact of any element of the system on $\Delta P_s$. 
In particular, an optimum of the transmission power of the source can be found and quantified. 
Moreover, a parametrization process can be defined, based on the analytical expression of $\Delta P_s$. Indeed, since 
$\Delta P_s$ 
characterizes the influence of the ambient backscatter parameters and the diffusing surface, 
it gives a precise knowledge of the contrast power variations, as a function of the variations of each parameter of the  backscattering system, and in particular the variations induced by moving any element. 
Furthermore, it is possible to control the value of the power contrast, by varying one or more parameters defining the ambient backscatter communication system. More particularly, the expression of $\Delta P_s$ makes it possible to search for one or more values of the system parameters for  maximizing the power contrast, or which make it possible to guarantee that the contrast power be greater than a power threshold, beyond which the decoding of a backscattered signal is ensured.
\begin{figure}[htbp]
\centering
\includegraphics[scale=0.45]{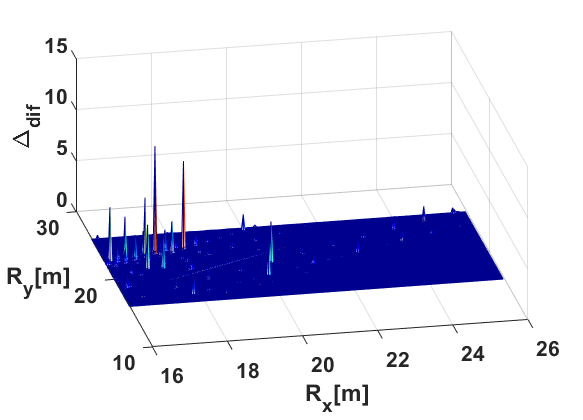}
\caption{\footnotesize Relative increase of the contrast $\Delta_{dif}$ due to the presence of the diffusing surface (3500 MHz) as a function of the position of the reader R.}
\label{Contrast3D}
\end{figure}

\begin{figure}[htbp]
\centering
\includegraphics[scale=0.45]{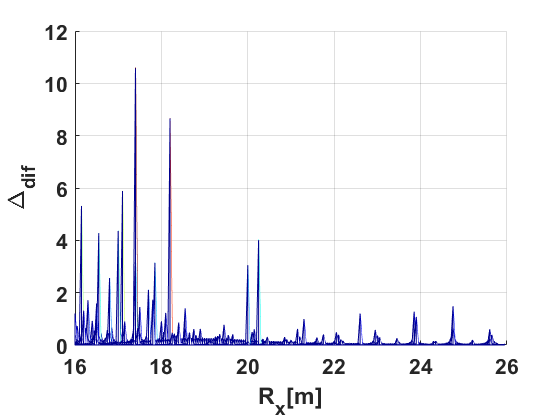}
\caption{\footnotesize Relative increase of the contrast $\Delta_{dif}$ due to the presence of the diffusing surface (3500 MHz), as a function of the coordinate $R_x$ of the reader R.}
\label{Contrast2D}
\end{figure}

The term  $P_{dif-tag} $
   of the contrast power (\ref{deltaP3}) shows the specific impact of the presence of the diffusing surface V. 
	This term may be positif or negative, depending on the configuration of the system, and particularly on the location of the reader. 
	Figures \ref{Contrast3D} and \ref{Contrast2D}, plotted for $P_{dif-tag} >0$ using numerical values (defined in Section \ref{sec:Numeric}), show the relative increase $\Delta_{dif}$ of the contrast power induced by the presence of surface V, compared to the case without V, by considering a set of potential positions of the reader (coordinates $R_x$ and $R_y$ for a reader R located in a plan). 
	These figures show that for a great number of locations of the reader, the presence of V increases the contrast power. Though this increase is relatively low for most potential locations of the reader (blue surface of Figure \ref{Contrast3D}, blue area close to $R_x$ axis of Figure \ref{Contrast2D}), it reaches high values for many locations (peaks of Figure \ref{Contrast3D} and Figure \ref{Contrast2D}). Moreover, the maximum increase reaches more than 10 times the contrast power value reached without the surface V.

\section{Application} \label{sec:application}

\subsection{Contrast to Noise Ratio and Performance} \label{sec:application1}
Let consider an energy detector, used by the reader to detect and decode the tag signal. 
The contrast to noise ratio ($\Delta_{SNR}$) between the two states
\begin{equation} \label{CNR}
\Delta_{SNR} =  \frac{\left|\sqrt{ P_1 + \Delta P_s}  - \sqrt{ P_1 }\right| }{\sqrt{N_{th}} } 
\end{equation}
makes it possible to determine the performance of such a detector, characterized by the BER (Bit Error Rate), 
by using the expression BER = 0.5 erfc($\Delta_{SNR}$) \cite{RachediPhan19}, where erfc(∙) is the complementary error function, $N_{th}$ is the noise average power, and SNR is the signal to noise ratio.

	\begin{figure}[htbp]
\centering
\includegraphics[scale=0.5]{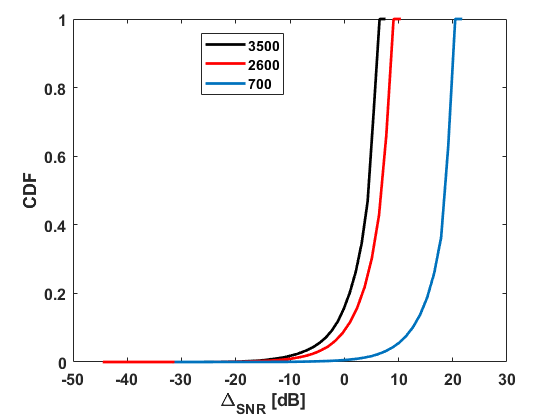}
\caption{\footnotesize CDF of the SNR variation $\Delta_{SNR}$ (dB), induced by the diffusing surface, for the frequencies 700 MHz, 2600 MHz, and 3500 MHz.}
\label{DeltaSNR1}
\end{figure}

\begin{figure}[htbp]
\centering
\includegraphics[scale=0.5]{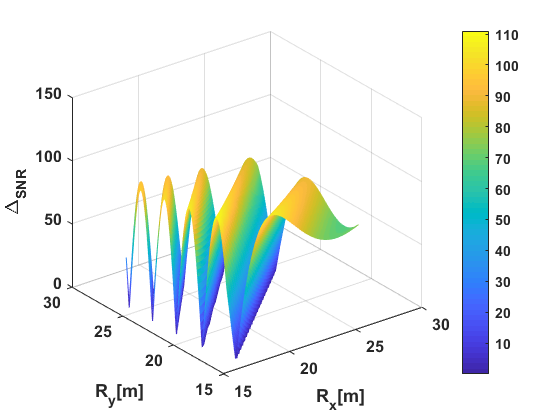}
\caption{\footnotesize Locations of the SNR variation $\Delta_{SNR}$, due to diffusing (700 MHz).}
\label{DeltaSNR2}
\end{figure}

\begin{figure}[htbp]
\centering
\includegraphics[scale=0.5]{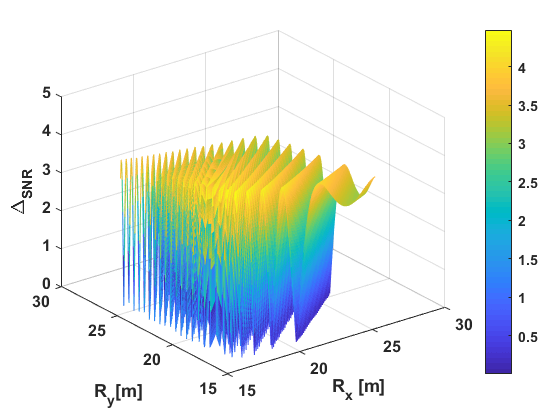}
\caption{\footnotesize Locations of the SNR variation $\Delta_{SNR}$, due to diffusing (3500 MHz).}
\label{DeltaSNR3}
\end{figure}

\subsubsection*{Numerical application} \label{sec:Numeric}
We consider the system (Figure 4).
The coordinates of each element of the system (expressed in meters) are 
S(0,0,0), A(6,6,0), B(8,6,0), T(-10,-10, 0). The 
set of potential locations of the reader is given by R($R_x$, $R_y$,0), where the range of 
variation of $R_x$ is [16,26], and the range of 
variation of $R_y$ is [16,26]; source power P =~30 dBm (e.g. WiFi source), antenna gains and diffusing coefficient $G^i$ =~1 
where i $\in$ (s, r, t, d), noise average power $N_{th}$ =~-116 dBm.

Figure \ref{DeltaSNR1} represents the cumulative distribution function (CDF) of $\Delta_{SNR}$, for the frequencies 700, 2600, and 3500 MHz.
It can be observed that this CDF is a decreasing function of the frequency. 
For example, for a value of the CDF equals to 0.2, 
$\Delta_{SNR}$ =15~dB (700 MHz) and 0~dB (3500 MHz). 
Moreover, for $\Delta_{SNR} = 3$~dB, the CDF reaches about 3\% (700 MHz). 
This means that for about 97\% of the possible locations of the reader considered in this system, the SNR is multiplied by a factor at least equal to 2. 
And the SNR is multiplied by a factor 
at least equal to 10 ($\Delta_{SNR} = 10$~dB) for about 95\% 
of the locations (CDF value equal to 0.05)
for the frequency 700 MHz. 
This figure also shows that for high frequency values (2600 and 3500 MHz), the improvement of the SNR induced by the  contrast power is relatively low.

Figures \ref{DeltaSNR2} and \ref{DeltaSNR3} represent maps of $\Delta_{SNR}$ values, according to the potential positions of the reader R in the plan. 
It can be observed in Figure \ref{DeltaSNR2} (700 MHz) that maxima values reach about a factor 110,
and the maxima values observed Figure \ref{DeltaSNR3} (3500 MHz) are relatively low, reaching about a factor 4.
Therefore, for many locations of the reader, the diffusing surface V induces a strong relative increase of the SNR. As a consequence, by locating the reader in given locations, the BER performance can be improved. 
The analysis developed makes it possible to determine with accuracy the positions where to locate the reader for reaching the best performance.

\subsection{Energetic Performance Optimization} \label{sec:application1}
Let consider the system need a power threshold value $P_{thr}$ to be able to decode the backscattered signal.
For a given set of parameters of the system, 
the minimum transmitting power value of the source must answer 
\begin{eqnarray} \label{signaldirect1}
\Delta P_s \geq  P_{thr}.
\end{eqnarray}

To reach the threshold value $P_{thr}$ in a configuration without the diffusing surface, the source S has to transmit a power P.
A power $P_s$ is needed in a configuration with the diffusing surface. 
%
Therefore the transmit power of the source has to answer the following criterion $ P  \geq  \frac{P_{thr}}{A_0}$ where
\begin{eqnarray} \label{P0}
A_{0} &=& {A'}_{tag}^2 + 2 {A'}_{dir} {A'}_{tag}  \cos⁡(k(r_t+r'_t-r)). \nonumber \\
\end{eqnarray}
and $ P_s  \geq   \frac{P_{thr}}{A_s} $
where
\begin{eqnarray} \label{P0}
A_{s} &=& {A'}_{tag}^2 + 2 {A'}_{dir} {A'}_{tag}  \cos⁡(k(r_t+r'_t+r)) \nonumber \\
      &+& 2 {A'}_{dif} {A'}_{tag} \sin⁡(\frac{ke}{2}) \nonumber \\
			&\times& \cos⁡[k(r_1+ r'_1-(r_t+r'_t)+\frac{e}{2})].  \nonumber \\
\end{eqnarray}
where  ${A'}_{dir} = \frac{A_{dir}}{P}$,  ${A'}_{tag} = \frac{A_{tag}}{P}$ and ${A'}_{dif} = \frac{A_{dif}}{P}$.
Therefore, the formula of $\Delta P_s$ makes it possible to reach the threshold contrast power value, by minimizing
the source transmit power. Indeed $\Delta P_s \geq 0$ and consequently $A_s \geq A_0$, for given characteristics of the system (position of the tag, the surface and the reader and carrier frequency), and in particular for many locations of the reader (see section \ref{sec:powercontrast}).
Therefore  $P_s \leq P$.
This makes it possible to quantify the amount of energy that can be saved, thanks to the diffusing surface. Moreover, this allows an optimization of the energetic performance, too. Some potential applications can be those powered by 5G base stations.
\section{Conclusion} \label{sec:conclusion} 
The analytical model developed in this article makes it possible to analyze an ambient backscatter communication system scenario, constituted by a transmit source, a retro-diffusing tag, a reader and a diffusing surface. The established power contrast closed form expression highlights the impact of the system configuration parameters. This expression allows a performance analysis in a simple and quick way. As an application, we show that the presence of the diffusing surface improves the power contrast, in many configurations of the system, and therefore the detection of the
message transmitted by the tag. Moreover, the contrast can be maximized, by considering appropriate location readers, determined thanks to the closed form formula. Furthermore, we show that the closed form expression makes it possible to analyze the BER performance, to quantify with accuracy the energetic performance and to optimize it. In future papers, we will extend this model to take into account other configurations, such as plane diffusing surfaces.



%

\appendix
For a low value of h, we can write (first order approximation) 
\begin{eqnarray} \label{signaldirect1}
 r_n &\approx& r_1+ \frac{ah}{r_1}   \nonumber  \\  	
 {r'}_n &\approx&  {r'}_1- \frac{bh}{{r'}_1}  \nonumber  \\ 
  r_{j+1} &\approx& r_1+ \frac{ajh}{n r_1}   \nonumber  \\  	
 {r'}_{j+1} &\approx&  {r'}_1 - \frac{bjh}{n{r'}_1}  \nonumber  \\ 
 r_{j+1} + {r'}_{j+1} &\approx& r_1 + {r'}_1 +j e', 	  \forall j = 0...n     \nonumber   \\
 r_1^2 {r'}_1^2 & \approx & r_j^2 {r'}_j^2,   \forall j = 1...n   \nonumber   \\     
r_n + {r'}_n &\approx& r_1 + {r'}_1 +n e' \nonumber   \\
 r_1^2 {r'}_1^2 & \approx & r_j^2 {r'}_j^2,   \forall j = 1...n 
\end{eqnarray}

\subsection*{Continuum Model Diffused Signal Expression } \label{powercalculationscontinuum}
We consider a continuous homogeneous diffusing surface. Each elementary part of the surface diffuses the power it receives. Therefore, by considering the expressions and approximations, 
we can express the discrete sum as an integral, as follows (considering that n+1 rays coming from the source are diffused by the surface):

\begin{eqnarray} \label{signaldifcalcul}
 Y_{dif}  &=&  \sum_{j=0}^{n} \sqrt{\frac{K^{str}P}{r_{j+1}^2}} \sqrt{\frac{1}{{r'}_{j+1}^2}} \exp i(wt-k(r_{j+1}+{r'}_{j+1})) \nonumber \\
				& \approx & \sqrt{\frac{K^{sar}P}{r_1^2 {r'}_1^2}} \exp i(wt)\exp -ik(r_1+{r'}_1) \nonumber \\
				& \times & \sum_{j=0}^n  \exp -i(k j e') \nonumber \\
				& \approx & \sqrt{\frac{K^{sar}P}{r_1^2 {r'}_1^2}} \exp i(wt)\exp -ik(r_1+{r'}_1)\nonumber \\
				& \times & \int_{0}^{h}  \exp -i(k \alpha u) du \nonumber \\
				& \approx & \sqrt{\frac{K^{sar}P}{r_1^2 {r'}_1^2}} e^{i(wt)} e^{-ik(r_1+{r'}_1)} \frac{e^{-i(k e)}-1 }{-ik \alpha}  \nonumber \\
					& \geq &  \sqrt{\frac{K^{sar}P}{r_1^2 {r'}_1^2}} e^{i(wt)} e^{-ik(r_1+{r'}_1)} \frac{1 - e^{-i(k e)} }{ik}	\nonumber \\
					& \geq & \frac{2}{k} \sqrt{\frac{K^{sar}P}{r_1^2 {r'}_1^2}} e^{i(wt)} e^{-ik(r_1+{r'}_1 +\frac{e}{2} )} \sin{\frac{k e}{2}},    \nonumber \\
\end{eqnarray}
for $\alpha \leq 1 $, where $\alpha = \frac{a}{r_1} - \frac{b}{{r'}_1}$, $e' =  \alpha  \frac{h}{n}$ and $e = \alpha h$. 
The signal at the reader, resulting of a diffusion by the surface V, reaches an amplitude at least equal to (\ref{signaldifcalcul2}):
\begin{eqnarray} \label{signaldifcalcul2}
 Y_{dif}  & = & \frac{2}{k} \sqrt{\frac{K^{str}P}{r_1^2 {r'}_1^2}} e^{i(wt)} e^{-ik(r_1+{r'}_1 +\frac{e}{2} )} \sin{\frac{k e}{2}}.    \nonumber \\
\end{eqnarray}


%


\begin{thebibliography}{1}

\bibitem{Liu13}
V. Liu, A. Parks, V. Talla, S. Gollakota, D. Wetherall, and J. R. Smith, “Ambient backscatter: wireless communication out of thin air,” in Proc. of ACM SIGCOMM 2013, Hong Kong, China, Aug. 2013, pp. 39-50.

\bibitem{Huynh18}
N. Van Huynh, D. T. Hoang, X. Lu, D. Niyato, P. Wang and D. I. Kim, "Ambient Backscatter Communications: A Contemporary Survey," in IEEE Communications Surveys and Tutorials, vol. 20, no. 4, pp. 2889-2922, Fourthquarter 2018.

\bibitem{RachediPhan19}
K. Rachedi, D.-T. Phan-Huy, N. Selmene, A. Ourir, M. Gautier, A. Gati, A. Galindo-Serrano, R. Fara, J. de Rosny “Demo Abstract : Real-Time Ambient Backscatter Demonstration,” in Proc. IEEE INFOCOM 2019, pp. 987–988, 2019.


\bibitem{Fara20}
R. Fara, D.-T. Phan-Huy, M. Di Renzo, “Ambient backscatters-friendly 5G networks: creating hot spots for tags and good spots for readers,” accepted to IEEE WCNC 2020, may 2020. 

\bibitem{Gati19}
A. Gati et al. “Key technologies to accelerate the ICT Green evolution - An operator's point of view” submitted to IEEE, 2019. Available at: https://arxiv.org/abs/1903.09627 

\bibitem{Wan16}
G. Wang, F. Gao, R. Fan and C. Tellambura, "Ambient Backscatter Communication Systems: Detection and Performance Analysis," in IEEE Transactions on Communications, vol. 64, no. 11, pp. 4836-4846, Nov. 2016.

\bibitem{Yan18}
G. Yang, Y. Liang, R. Zhang and Y. Pei, "Modulation in the Air: Backscatter Communication Over Ambient OFDM Carrier," in IEEE Transactions on Communications, vol. 66, no. 3, pp. 1219-1233, March 2018.

\bibitem{Devi19}
J. K. Devineni and H. S. Dhillon, “Non-coherent detection and bit error rate for an ambient backscatter link in time-selective fading,” arXiv preprint arXiv:1908.05657, 2019, available at: https://arxiv.org/pdf/1908.05657.pdf.



\bibitem{ITU15}
Recommendation ITU-R P.2040-1 (07/2015) ”Effects of building materials and structures on radiowave propagation above about 100 MHz-P Series-Radiowave”.




\bibitem{Renzo19}
Renzo, M.D et al. « Smart radio environments empowered by reconfigurable AI meta-surfaces: an idea whose time has come.” J Wireless Com Network 2019, 129 (2019).


\bibitem{Devi18}
J. K. Devineni and H. S. Dhillon, "Exact Bit Error Rate Analysis of Ambient Backscatter Systems Under Fading Channels," 2018 IEEE 88th Vehicular Technology Conference (VTC-Fall), Chicago, IL, USA, 2018, pp. 1-6.






\bibitem{Grffin09}
J. Griffin, ”High-frequency modulated-backscatter communication using multiple antennas”, 2009.Available at: https://smartech.gatech.edu/handle/1853/28087.
\end{thebibliography}
\end{document}